# Online Dissolved Gas Analysis (DGA) Monitoring System

Xianda, Deng

University of Tennessee, Knoxville
United States of America

Kyle Thomas, Huiying Huang,
Scott P Adams, Hesen Liu.
Dominion Energy, Inc.
United States of America

**SUMMARY**

Transformers are critical assets in power systems and transformer failures can cause asset damage, customer outages, and safety concerns. Dominion Energy has a sophisticated monitoring process for the transformers. One of the most cost-efficient, convenient and practical transformer monitoring methods in industry is Dissolved Gas Analysis(DGA). Leveraging new technology, on-line transformer monitoring equipment is able to measure samples automatically. The challenges of unstable sampling measurements and contradicted analysis results for DGA are discussed in this paper. To provide further insight of transformer health and support a new transformer monitoring process in Dominion Energy, a DGA monitoring system is proposed. The DGA analysis methods used in the monitoring system are selected based on laboratory verification results from Dominion Energy. After derive the thresholds from IEEE standard, the solution of the proposed monitoring system and test results are presented. In the end, a historical transformer failure case in Dominion was analyzed and the results indicate the monitoring system can provide prescient information and sufficient supplemental report for making operational decisions.

**KEYWORDS**

Power systems, Monitoring, Situation awareness, Transformer, online monitoring, DGA,

## I. Introduction

### a. *Background*

Situation awareness provides critical information for power systems operation and reliability; enormous efforts have been invested in monitoring power systems and their component status [1]-[7]. Transformers are some of the most essential and expensive assets in power transmission and distribution systems [8], [9]. Maintaining transformer reliability is critical for transmission and distribution system operations. Both malfunctions and normal operations of a transformer may generate certain types of gases and dissolved in the insulation system of the transformer. The combustible and non-combustible gases are excellent indicators of incipient fault conditions that may lead to failure. The most important diagnostic method to provide insight to the health of internal transformer components is Dissolved Gas Analysis (DGA) [10], [11]. In conventional process of DGA, transform oil samples are manually collected and delivered to a laboratory. The oil samples are analyzed by different methods with special equipment. As technology evolves, commercial on-line measurement devices are available to collect immediate measurements of dissolved gases in transformers without human effort. This provides an opportunity to automate the DGA process and develop on-line DGA monitoring systems.

A sophisticated transformer monitoring process has been employed by Dominion Energy for years. In recent years, on-line measuring equipment has been installed in the field and sending the measurements to a laboratory for evaluation. The comparisons between analysis results of on-line measurements and manually collected measurements indicate that the on-line measurements have a high accuracy, while requiring less human labor. To facilitate on-line measurements, a DGA monitoring system was implemented and applied in the Dominion Energy transformer monitoring process. The monitoring system evaluates DGA results from the laboratory daily and determines the health status of transformers based on predetermined gas detection levels. The system will inform the operation team and engineers once any gas is out of normal status ranges.

### b. *Motivation and challenges*

Recent transformer monitoring events show that gases concentration and trends may change dramatically in certain failure scenarios. In those extreme situations, cascading failure of transformers may occur in a short period of time. In one of recent transformer failure, both of gas concentration and concentration trending kept increasing in one day until transformer failure happened. The scene of the transformer failure is shown in Fig. 1. For personnel safety and equipment reliability purpose, the present process needs to be improved in order to help engineers make timely operational decisions before a transformer fails. An on-line DGA monitoring system, which provides essential information and timely notifications, is desired for supporting the new transformer monitoring process.

One of the challenges for online DGA analysis is that DGA sampling measurements and analysis are not always stable. In some cases, DGA analysis results contradict each other [12]. It is difficult to select one single reliable DGA for the transformer monitoring process. On the other side, using multiple DGA analysis results may lead to confusion when making operational decisions.

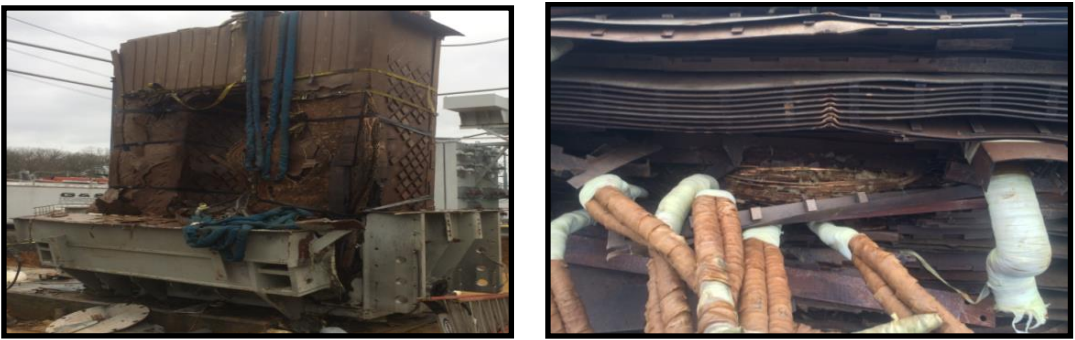

Figure 1. Transformer failure in a recent event

In this paper, the implementation of an online DGA monitoring system is presented and this paper is organized as follows: In section II, three DGA methods and the rules of the methods in the monitoring system are discussed. The system architecture and monitoring process are presented in section III. In section IV, the monitoring system is tested with simulated data and validated by one historical transformer failure case. The conclusion is drawn in section V.

**II. Methodology**

DGA is one of the most widely used and effective methods for transformer preventive maintenance in the industry. Multiple types of DGA methods are well defined in IEC and IEEE standard for industry users [13], [14]. The methods analyze gases in transformer insulate oil and identify the types of fault inside a transformer. Each type of DGA method may evaluate different types of gas for analysis. In Dominion Energy online DGA monitoring system, the Key Gas method is employed as the primary analysis method for event thresholds. Ratio methods and Total Dissolved Combustible Gas (TDCG) method are applied for supporting operational decisions. The details of these methods are listed below:

*a. Key Gas method*

In Dominion Energy DGA analysis process, only $C_2H_2$, $C_2H_4$, and $H_2$ are adopted due to the issues of DGA discussed before. Based on years of laboratory evaluation results between on-line samples and manual samples, these three gases provide a high degree of confidence for monitoring process. Three-level thresholds were derived from IEEE C57.104 for each type of the gases and listed in Table 1 and 2.

TABLE 1. Dominion Energy DGA concentration limits

| Gases | Level 1 (PPM) | Level 2 (PPM) | Level 3 (PPM) |
|---|---|---|---|
| $C_2H_2$ | 3.33 | 6.66 | 10 |
| $C_2H_4$ | 66.6 | 133.33 | 200 |
| $H_2$ | 333.3 | 666.6 | 1000 |

TABLE 2. Dominion Energy DGA concentration trend limits

| Gases | Level 1 (PPM/day) | Level 2 (PPM/day) | Level 3 (PPM/day) |
|---|---|---|---|
| $C_2H_2$ | 1.66 | 3.33 | 5 |
| $C_2H_4$ | 3.33 | 6.66 | 10 |
| $H_2$ | 8.33 | 16.66 | 25 |

b. Ratio methods

Ratio methods use different combustible gases to diagnose one single potential fault in transformers [13]. The process of interoperating the ratio methods is highly empirical and depends on the experience of investors. The methods have been validated in European and partially approved in the U.S.. Although the ratio methods may not provide high confidence analysis results for monitoring process, it may be considered as supplementary information for making operational decisions. In [14], five types of ratio methods base on thermal degradation principles are defined:

$$\text{Ratio 1 (R1)} = CH_4 / H_2 \quad (1)$$
$$\text{Ratio 2 (R2)} = C_2H_2 / C_2H_4 \quad (2)$$
$$\text{Ratio 3 (R3)} = C_2H_2 / CH_4 \quad (3)$$
$$\text{Ratio 4 (R4)} = C_2H_6 / C_2H_2 \quad (4)$$
$$\text{Ratio 5 (R5)} = C_2H_4 / C_2H_6 \quad (5)$$

c. TDCG method

The TDCG method is a well-accepted technique to monitor transformer insulation failure inside transformers. It uses the total volume of the dissolved combustible gas as an indicator of potential faults. This method does not provided any criteria for determining fault conditions, but it can be applied constantly after incipient fault condition deteriorates. The definition of the TDCG method is given below:

$$TDCGV = H_2 + CH_4 + C_2H_6 + C_2H_4 + C_2H_2 + CO \quad (6)$$

Where the unit of $H_2$, $CH_4$, $C_2H_6$, $C_2H_4$, $C_2H_2$ and $CO$ is liter/liter (ppm)

III. Solution and implementation

The transmission transformers fleet in Dominion Energy are currently modeled in a real-time data management system, PI. The field measurements from DGA on-line sampling equipments are streaming and stored in the PI historian. To fully utilize the present implementation in Dominion Energy, the proposed DGA monitoring system employs the PI system as infrastructure.

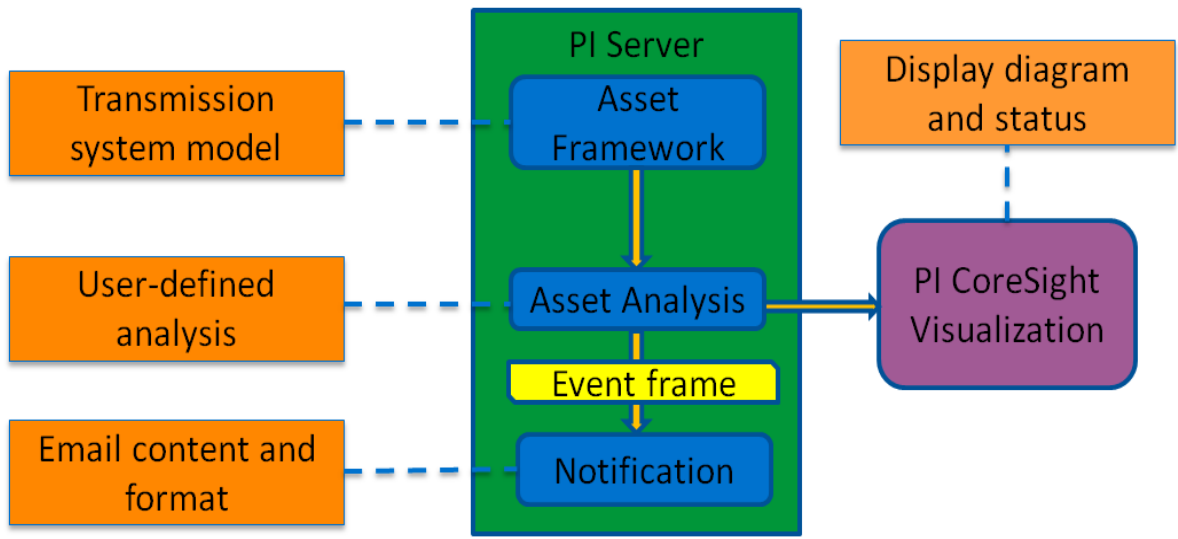

Figure 2. Pi system infrastructure for the proposed monitoring system

The PI system provides multiple software packages and tools for data management, analysis, visualization, client application developments, etc [15]. In the online DGA monitoring system, the following components from the PI system are adopted as shown in Fig. 2: (1). Asset Framework, (2). Asset analysis service, (3). Notification service and (4). PI Coresight. PI Asset Framework is a repository, where users define asset models, hierarchies and references for both PI and no-PI data sources. The models and references defined in the PI Asset Framework are used as the data sources in other components. In Asset analysis service, users can define their own process to perform calculation and analysis for decision making. Users can also define the conditions that will trigger an event. PI Notification service is designed to send a notification email when a pre-defined event is detected in the PI database. It allows users to configure the notification email template and recipients. Both data and analysis results can be included in the notification email. The PI Coresight is a web-based visualization tool with Windows Integrated Security. Users can easily design and build their own graphic user interface with multiple formats. The data and event information in the PI database are automatically displayed at the user-defined interface.

The system architecture of the online DGA monitoring system is shown in Fig. 3. The transformer information, available gas measurements and the thresholds in Table 1 and 2 are modeled in PI Asset Framework. Key gas methods are implemented in the Asset analysis server for primary monitoring process and event generation analysis are implemented for each threshold. Ratio methods and TDCG method are implemented and used as supplementary information to support operational decisions.

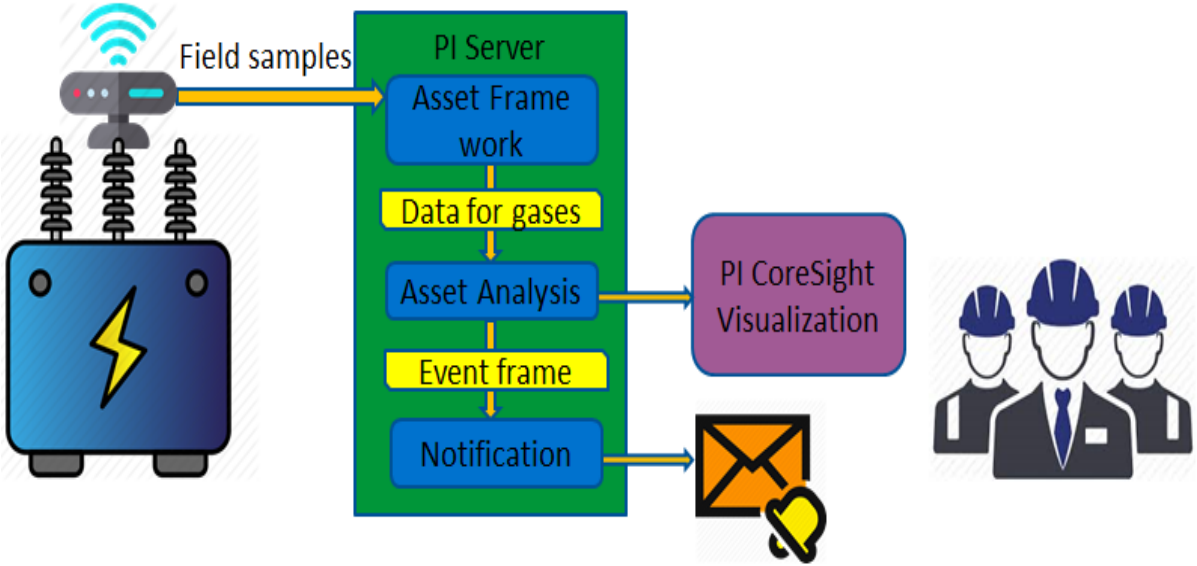

Figure 3. Online DGA Monitoring System Architecture

The flow chart of the monitoring process is shown in Figure 4. The analysis process evaluates the gas concentrations and concentration trends when new measurements are stored in the PI database. The status of gas concentrations and concentration trends will be determined based on the thresholds. If the measurement is over any threshold, an event will be generated and the monitoring process will determine event severity based on a DGA severity of fault table from IEEE standard as shown in Fig. 5. The determined status and analysis results are stored in the event frame for visualization and notification. An example of notification email and visualization from a simulated case is shown in Figure 6 and 7.

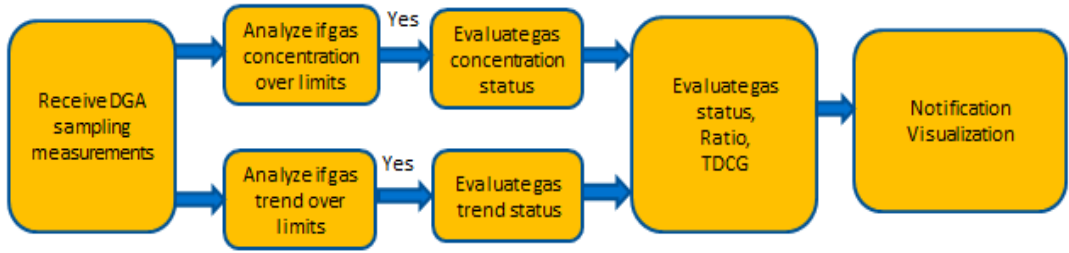

Figure 4. Flow chart of the DGA analysis process

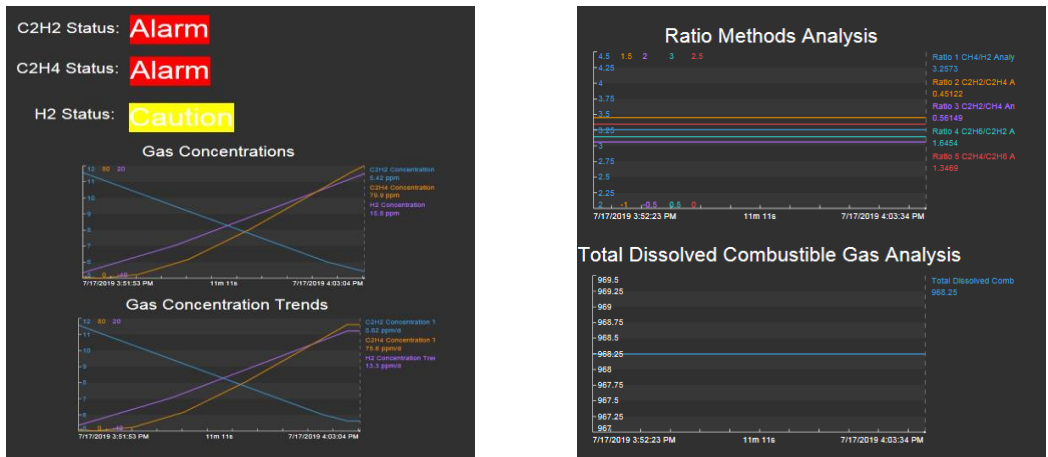

Figure 5. DGA severity of fault based on IEEE C57.104 standard

Figure 6. An email notification sample generated from a test event

Figure 7. PI Coresight display of online DGA monitoring system

## IV. Test and Case analysis

To further validate the online DGA system, a transformer failure in Dominion Energy grid was analyzed. In the transformer, only $H_2$ concentration and concentration trend were sampled for the online DGA monitoring process. In the case, $H_2$ concentration and concentration trend of the transformer started to increase one day before the failure as shown in Fig. 8 and 9. However, the old DGA monitoring system only sends out one notification per day. In this case, a notification was sent three hours before the failure and technicians had limited time to examine the transformer or have sufficient information to make an operational decision.

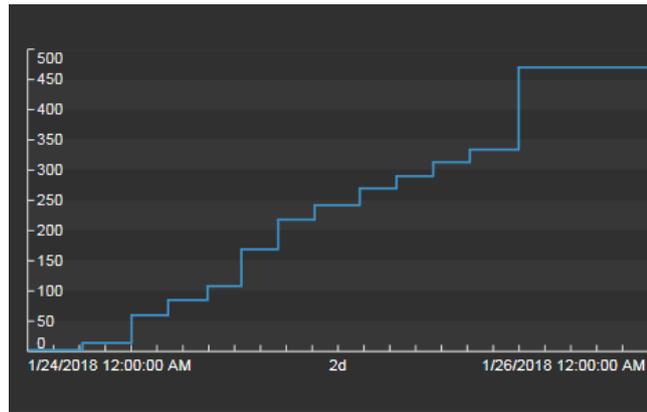

Figure 8. TX-1C H2 concentration level during a transformer failure

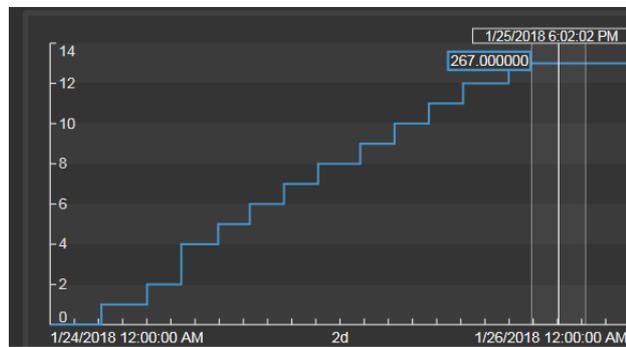

Figure 9. TX-1C H2 concentration trend during a transformer failure

The test case is archived in the Dominion Energy PI database and the historical data was backfilled into the database for the test. The online DGA monitoring system automatically retrieved the data and performed DGA analysis. Three event frames and notification emails were generated and a summary of the events are listed in Table 3. As shown in Fig. 8 and 9, $H_2$ concentration trend was higher than threshold level 2 at 08:02 AM, Jan 24th, while the concentration was below threshold level 1. The gas status was recognized as cautious and the first event was generated and an email was sent out. About 3 hours later, $H_2$ concentration trend was higher than threshold level 3 and gas status was determined as cautious. The DGA monitoring system sent out the second email. On Jan 25th morning, $H_2$ concentration exceeded level 1 while its trend was still above threshold level 3. The gas status was determined as warning. The third notification was triggered 3 hours before the failure. As demonstrated in this case, the online-DGA monitoring system provides more notifications for operators to conduct necessary actions when an internal fault happens in a transformer. The multiple-level thresholds provide more prescient transformer health information.

TABLE 3. A summary of events in a transformer failure case

| Time | H2 concentration (PPM) | H2 Trend (PPM/day) | Time before failure |
| --- | --- | --- | --- |
| 1/24 00:00 | 2 | 0 | ≈ 37 hours |
| 1/24 08:02 | 60 | 16 | ≈ 29 hours |
| 1/24 10:55 | 85 | 31 | ≈ 26 hours |
| 1/25 10:14 | 334 | 228 | ≈ 3 hours |

## V. Conclusion

To improve the transformer monitoring process in Dominion Energy and leverage new technology, an online DGA monitoring system was implemented based on PI system. Multiple-level thresholds derived from IEEE standard were imported in the system. The online DGA monitoring system and the proposed thresholds were tested with simulated data and validated by historical data from an actual transformer failure. The test results indicate that the online DGA monitoring system is able to detect dissolved gas changes efficiently, provides supplement DGA analysis information for making operational decision. The notifications and visualizations in the online DGA monitoring system guarantee that the operators will not miss any important gas status changes in transformers.